\documentclass[sigconf]{acmart}

\AtBeginDocument{%
  \providecommand\BibTeX{{%
    \normalfont B\kern-0.5em{\scshape i\kern-0.25em b}\kern-0.8em\TeX}}}

\copyrightyear{2023} 
\acmYear{2023} 
\setcopyright{acmlicensed}\acmConference[MM '23]{Proceedings of the 31st ACM International Conference on Multimedia}{October 29-November 3, 2023}{Ottawa, ON, Canada}
\acmBooktitle{Proceedings of the 31st ACM International Conference on Multimedia (MM '23), October 29-November 3, 2023, Ottawa, ON, Canada}
\acmPrice{15.00}
\acmDOI{10.1145/3581783.3611704}
\acmISBN{979-8-4007-0108-5/23/10}

\usepackage{xcolor}
\usepackage{color}
\usepackage{booktabs}
\usepackage{amsmath}
\usepackage{graphicx}
\usepackage{xspace}
\usepackage{newtxmath}
\usepackage{rotating}
\usepackage{multirow}
\usepackage{url}
\usepackage[linesnumbered, ruled]{algorithm2e}
\usepackage{amsmath}
\usepackage{makecell}
\usepackage{balance}
\usepackage{subcaption}
\usepackage{cleveref}
\usepackage{subcaption,multirow,array,dsfont,ctable,xcolor}
\usepackage{diagbox}
\usepackage{bbding}
\newlength\savewidth\newcommand\shline{\noalign{\global\savewidth\arrayrulewidth
  \global\arrayrulewidth 1.5pt}\hline\noalign{\global\arrayrulewidth\savewidth}}
\newcommand{\methodname}{Emo-DNA\xspace}
\newcommand{\prototypeloss}{decouple}
\newcommand{\alignmentloss}{align}
\newcommand{\contrastiveloss}{SCL\xspace}

\usepackage{enumitem}

\newcommand{\vct}[1]{\boldsymbol{#1}} % vector
\newcommand{\mat}[1]{\boldsymbol{#1}} % matrix
\definecolor{maroon}{rgb}{0,0,0}
\newcommand{\best}[1]{\textcolor{maroon}{\textbf{#1}}}
\newcommand{\etal}{\textit{et al}.\xspace}
\newcommand{\ie}{\textit{i}.\textit{e}.,\xspace}
\newcommand{\eg}{\textit{e}.\textit{g}.,\xspace}

\newcommand{\std}[1]{\scriptsize{#1}}
\makeatletter

\begin{document}
\title{Emo-DNA: Emotion Decoupling and Alignment Learning for Cross-Corpus Speech Emotion Recognition}

% Single author syntax
\author{Jiaxin Ye}
\affiliation{%
  \institution{ISTBI, Fudan University}
  \country{}}
\email{jxye22@m.fudan.edu.cn}

\author{Yujie Wei}
\affiliation{%
  \institution{ISTBI, Fudan University}
  \country{}}
\email{yjwei22@m.fudan.edu.cn}

\author{Xin-Cheng Wen}
\affiliation{%
  \institution{Harbin Institute of Technology}
  \country{}}
\email{xiamenwxc@foxmail.com}
  
\author{Chenglong Ma}
\affiliation{%
  \institution{ISTBI, Fudan University}
  \country{}}
\email{clma22@m.fudan.edu.cn}
  
\author{Zhizhong Huang}
\affiliation{%
  \institution{School of CS, Fudan University}
  \country{}}
\email{zzhuang19@fudan.edu.cn}

\author{Kunhong Liu}
\affiliation{%
  \institution{School of Film, Xiamen University}
  \country{}}
\email{lkhqz@xmu.edu.cn}

\author{Hongming Shan}
\authornote{Corresponding author}
\affiliation{%
  \institution{ISTBI \& 
MOE FCBS, Fudan University}
  \country{}}
\email{hmshan@fudan.edu.cn}  

\renewcommand{\shortauthors}{Jiaxin Ye et al.}

\begin{abstract}

Cross-corpus speech emotion recognition (SER) seeks to generalize the ability of inferring speech emotion from a well-labeled corpus to an unlabeled one, which is a rather challenging task due to the significant discrepancy between two corpora. 
Existing methods, typically based on unsupervised domain adaptation (UDA), struggle to learn corpus-invariant features by global distribution alignment, but unfortunately, the resulting features are mixed with corpus-specific features or not class-discriminative.
To tackle these challenges, we propose a novel Emotion Decoupling aNd Alignment learning framework (\methodname) for cross-corpus SER, a novel UDA method to learn emotion-relevant corpus-invariant features. 
The novelties of \methodname are two-fold: contrastive emotion decoupling and dual-level emotion alignment. 
On one hand, our contrastive emotion decoupling achieves decoupling learning via a contrastive decoupling loss to strengthen the separability of emotion-relevant features from corpus-specific ones.
On the other hand, our dual-level emotion alignment introduces an adaptive threshold pseudo-labeling to select confident target samples for class-level alignment, and performs corpus-level alignment to jointly guide model for learning class-discriminative corpus-invariant features across corpora.
Extensive experimental results demonstrate the superior performance of \methodname over the state-of-the-art methods in several cross-corpus scenarios.  Source code is available at \url{https://github.com/Jiaxin-Ye/Emo-DNA}.
\end{abstract}




\maketitle

\section{Introduction}
\label{Introduction}

Speech emotion recognition (SER) aims to automatically recognize human emotions from speech signals~\cite{schuller2018speech}, which has attracted much attention in human-computer interaction (HCI), mental diagnostic tools, in-car board systems, etc~\cite{Review_SER,wu2022automatic}. However, in these practical applications, significant discrepancies exist between training and real-world corpora, arising from different languages, cultures, speakers, contents, data scales, etc. These discrepancies across corpora lead to significant idiosyncratic variations impeding the generalization of current SER technology. Therefore, SER faces enormous challenges when being applied to cross-corpus scenarios, which require a strong generalization of inferring speech emotion from a well-labeled corpus to an unlabeled one.

To solve these challenges, many researchers have explored unsupervised domain adaptation (UDA)~\cite{CPAC_IJCAI,DANN_ICASSP,MM22_CSER} to match global distributions of source and target data, which aims to learn domain-invariant representations.
Recent advances in UDA have proven effective in addressing the challenge of corpus\footnote{Here, we use the term ``\emph{corpus}'' to refer to the ``\emph{domain}'' in the UDA. They are exchangeable throughout this paper.} misalignment for the cross-corpus SER~\cite{mismatch_SER}. 
One key idea to achieve corpus alignment is to reduce distribution shifts between the source and target corpus, which is commonly implemented through statistic divergence alignment and adversarial learning~\cite{UDA_review}.
Statistic divergence alignment methods~\cite{SER_MMD,SER_MKMMD} utilize various divergence measures, such as maximum mean discrepancy (MMD)~\cite{MMD}, to minimize domain discrepancy in a latent feature space. Lu~\etal~\cite{SER_MKMMD} propose a hierarchical alignment framework based on Multi-Kernel MMD (MK-MMD) criterion to learn corpus-invariant features.  
Instead of introducing statistic divergence measures, adversarial learning methods~\cite{ADDI_SOTA,DANN_ICASSP,SER_ICASSP_GRL} adaptively learn a measure of divergence under the guidance of adversarial domain discriminators. Feng~\etal~\cite{SER_ICASSP_GRL} present a noise representation learning framework that leveraged adversarial training to retain emotion-relevant information.

While these dominant UDA methods can effectively align the distributions across corpora at feature level and improve the generalization ability of models to some extent, they still suffer from the following two dilemmas in cross-corpus SER.
1) \textbf{False alignment} refers to falsely aligning features of different attributes (\eg emotion, corpus, language, and speaker) during domain adaptation, leading to performance degradation. Due to the nonlinear manifold structures underlying data distributions~\cite{IJCAI_disentangle}, the emotion-relevant and corpus-specific features might be highly entangled in corpus-invariant feature space. Most previous efforts focus on matching global marginal distributions between source and target corpora~\cite{class_confusion_DA}, which fail to decouple corpus information and preserve emotion information from the aligned distribution. This may lead to the false alignment problem, i.e., aligning emotion-irrelevant features (\eg corpus-specific ones) with emotion-relevant ones\footnote{Here, we use the term ``emotion-relevant features'' to refer to the features are emotion-discriminative. The ``corpus-specific features'' denotes features that contain specific information related to the corpus itself and are corpus-discriminative. }.
2) \textbf{Class confusion} refers to ignoring class consistency when aligning data distributions of different domains, leading to performance degradation. Although matching global source and target data distributions can pull the features of source and target corpora closer, it also mixes features of different classes. This can cause class confusion since the class-discriminative information from the source corpus cannot be easily transferred to the target corpus. In other words, if the model fails to maintain class consistency during corpus alignment, it ultimately results in the erroneous mapping of target corpus features (\ie negative transfer).

In this paper, we propose a novel \textbf{Emo}tion \textbf{D}ecoupling a\textbf{N}d \textbf{A}lignment learning framework (\textbf{\methodname}) to tackle the aforementioned issues for cross-corpus SER. For \methodname, we hope that feature decoupling and alignment are two key techniques to cross-corpus SER, like the complementary paired duplexes in deoxyribonucleic acid. 
On one hand, for contrastive emotion decoupling, we first leverage two encoders to learn emotion-relevant and corpus-specific features of each corpus, guided by source emotion labels and corpora labels, respectively. Then we propose a novel contrastive decoupling loss based on emotion-relevant and corpus-specific prototypes, which encourages decoupling emotion features from corpus-specific features by pushing emotion prototypes away from the corpus-specific ones. 
On the other hand, dual-level emotion alignment is based on class-level and corpus-level to achieve effective alignment. We leverage an adaptive threshold pseudo-labeling to select confident samples of the target corpus, and then devise a contrastive class alignment loss based on pseudo-labeled target samples and the source samples to encourage class alignment across corpora by maximizing the mutual information of representations of the same class. We further introduce an explicit corpus alignment loss to minimize corpus discrepancy across corpora considering the significant domain shifts between different corpora. 
During the training process, the contrastive emotion decoupling enforces the features to be emotion-relevant for the corpus alignment, and dual-level emotion alignment encourages the features to be emotion-relevant corpus-invariant with class discrimination. Decoupling and alignment promote progressive learning with each other and enable the model to achieve robust adaptability in the target corpus, establishing new state-of-the-art performance on several benchmark datasets. 

The main contributions of this paper are summarized as follows:
\begin{enumerate}[topsep=0pt]
\item[1)] We propose a novel UDA framework for cross-corpus SER,  called \methodname, which decouples emotion-relevant features from highly coupled distorted feature space and learns the emotion-relevant corpus-invariant features. To the best of our knowledge, \methodname makes the first attempt at identifying the false alignment and class confusion problems as the key limiting factors for cross-corpus SER.
\item[2)] We propose a contrastive emotion decoupling module to strengthen the separability of emotion-relevant features from corpus-specific ones in a contrastive learning framework.
\item[3)] We propose a dual-level emotion alignment module based on the class- and corpus-level to jointly guide the model to learn class-discriminative corpus-invariant features.
\item[4)] Extensive experimental results demonstrate effectiveness and superiority of \methodname over the state-of-the-art methods in several cross-corpus scenarios.
\end{enumerate}

\begin{figure*}[t]
	\centering
	\includegraphics[width=0.96\textwidth]{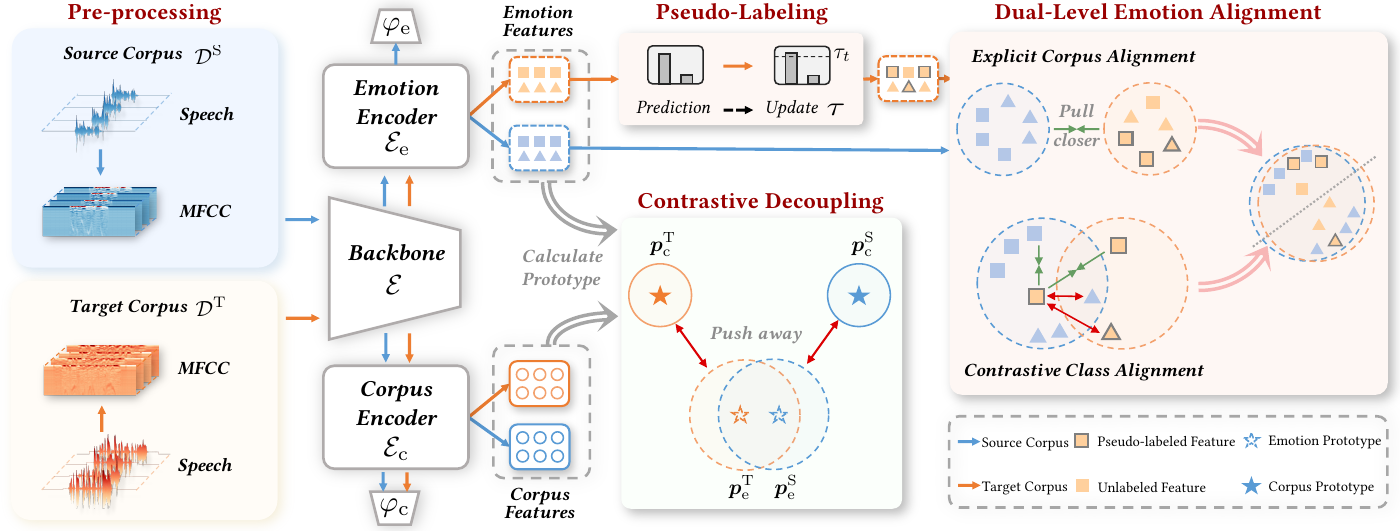}
	\caption{The overview of the proposed Emotion Decoupling aNd Alignment (\methodname). We use different colours to indicate the flow direction of different input corpora. For each corpus, we use two encoders with a shared backbone to learn emotion and corpus features, then calculate their prototype for contrastive decoupling. The pseudo-labeling module selects confident samples of the target corpus according to the estimated threshold $\tau_t$. Finally, the dual-level emotion alignment module conducts the explicit corpus alignment and contrastive class alignment to learn emotion-relevant corpus-invariant features. Note that the feature spaces shown in contrastive decoupling and dual-level emotion alignment are identical.}
	\label{fig:architecture}
\end{figure*}

\section{Related Works}
\label{Relatedworks}

\noindent\textbf{Unsupervised domain adaptation.}\quad The unsupervised domain adaptation (UDA) attempts to make the models generalize well from the source and target domains. Most previous works in cross-corpus SER aim to transfer the ability of inferring speech emotion from a labeled source corpus to an unlabeled target corpus. These methods can be summarized into two categories: statistic divergence alignment and adversarial learning. Statistic divergence alignment methods align distributions by minimizing domain discrepancy measurement. Recent UDA approaches minimize some measurements of corpus shift, such as Maximum Mean Discrepancy (MMD)~\cite{SER_MMD} and multi-kernel MMD~\cite{SER_MKMMD}. 
Another predominant approach to UDA is adversarial learning, which uses a domain discriminator that distinguishes whether the features are from the source or target corpus to encourage learning corpus-invariant feature in adversarial manner~\cite{CPAC_IJCAI,SER_ICASSP_GRL,DANN_ICASSP}. 
For example, Wen~\etal~\cite{CPAC_IJCAI} propose an adversarial training framework with the margin disparity discrepancy loss to align feature distribution. Feng~\etal~\cite{SER_ICASSP_GRL} present a noise representation learning architecture with adversarial training to keep emotion-relevant information. 
These explicit feature alignment methods fail to decouple emotion-relevant features from the highly entangled feature space and maintain class consistency, potentially resulting in false alignment and class confusion problems. Alternatively, we perform contrastive emotion decoupling and dual-level emotion alignment to alleviate these issues.

\noindent\textbf{Decoupled representation learning.}\quad The core idea of decoupled representation learning follows a prior that the model can learn to embed representations in a corpus-specific information space and a corpus-invariant emotion space. It allows the emotion information available for both corpora by assuming a shared corpus-invariant space. Recently, Tjandra~\etal~\cite{style_speech_disentangle} present unsupervised learning for speech synthesis by decoupling contents and style of speech representation. Xi~\etal~\cite{single_SER_disentangle} propose frontend attributes decoupling for single-corpus SER. 
However, these methods still focus on the individual corpus tasks rather than improving generalizability across corpora. 
In contrast to previous approaches, we explore the decoupled representation learning for UDA, which can enhance generalizability under different adaptation scenarios.

\noindent\textbf{Contrastive learning.}\quad
Contrastive learning~\cite{MoCo,SimCLR,BYOL} aims to learn generalized features by leveraging positive and negative views of samples. Unfortunately, it is challenging to apply contrastive learning directly in UDA due to the domain gap and unavailable class labels of the target domain.
To address this issue, many UDA methods~\cite{MM_contrastive_UDA,CVPR_contrastive_UDA,MM_contrastive_UDA2} combine contrastive learning with pseudo-labeling techniques to implicitly reduce the domain discrepancy. 
Ma~\etal~\cite{MM_contrastive_UDA2} propose a contrastive-based alignment strategy on two domain-oriented feature spaces with a target pseudo-label refinement.
Chen~\etal~\cite{MM_contrastive_UDA} integrate self-supervised pretext tasks and UDA in a contrastive manner to improve generalizability of the conventional self-supervised frameworks. 
% can not effectively reduce domain shifts and 
However, these implicit alignment methods do not guarantee that the aligned features via contrastive learning are only relevant to the semantic information (\ie emotion-relevant information), leading to the false alignment. Instead, we propose the contrastive emotion decoupling to learn emotion-relevant features decoupled from corpus-specific features.

\section{METHOD}
\label{Methodology}

\subsection{Problem Definition}

Given a source corpus $\mathcal{D}^\text{S}=\{(\vct{X}_i^{\text{S}}, {y}_i^{\text{S}})\}_{i=1}^{n^\text{S}}$ with $n^\text{S}$ well-labeled samples and an unlabeled target corpus $\mathcal{D}^\text{T}=\{\vct{X}_j^{\text{T}}\}_{j=1}^{n^\text{T}}$ with $n^\text{T}$ unlabeled samples, where $\vct{X}^\text{S}$, ${y}^\text{S}$ and $\vct{X}^\text{T}$ denote the source speech signals (or features), source emotion label and target speech signals (or features), respectively, the goal of cross-corpus SER is to learn an emotion classifier $\varphi_\text{e}$ that is trained on the source corpus $\mathcal{D}^\text{S}$ with source labels and minimizes the empirical risk on the target corpus $\mathcal{D}^\text{T}$. That is, the cross-corpus SER task aims to generalize the ability of inferring speech emotion from the labeled source corpus to the unlabeled target corpus, which is a typical unsupervised domain adaption problem.  In this paper, $\vct{X}^\text{S}$ and $\vct{X}^\text{T}$ denote the pre-processed features, and the source corpus and target corpus are characterized by probability distributions $P$ and $Q$, respectively.

\subsection{Overview of \methodname}

Figure~\ref{fig:architecture} presents the overview of the proposed \methodname framework  for cross-corpus SER. 
In a nutshell, \methodname performs contrastive emotion decoupling to learn decoupled emotion-relevant features and the dual-level emotion alignment to further learn class-discriminative corpus-invariant features.
Specifically, \methodname alleviates the false alignment and class confusion problem in two ways: 1) {\emph{Contrastive emotion decoupling}}: the emotion encoder $\mathcal{E}_\text{e}$ and corpus encoder $\mathcal{E}_\text{c}$ first learn emotion-relevant and corpus-specific features, respectively, and then the contrastive decoupling loss guides model to decouple emotion-relevant features from corpus-specific features by pushing emotion-relevant prototypes away from the corpus-specific prototypes. 
2) {\emph{Dual-level emotion alignment}}: the adaptive threshold pseudo-labeling first selects confident samples of the target corpus, and the contrastive class alignment loss based on pseudo-labeled target samples and the source samples encourages model to align features by maximizing the mutual information of representations sharing the same emotion. Moreover, the explicit corpus alignment loss aims to reduce domain shifts between two corpora by minimizing MK-MMD measure across corpora.
In the following subsections, we describe each of the proposed components in detail.

\subsection{Contrastive Emotion Decoupling}

\noindent{\textbf{Feature decoupling.}}\quad 
To transfer knowledge effectively, it is crucial for the model to learn useful emotion representations from pre-processed Mel-Frequency Cepstral Coefficients (MFCCs) features. 
As shown in Figure~\ref{fig:architecture}, labeled samples in $\mathcal{D}^\text{S}$ and unlabeled samples in $\mathcal{D}^\text{T}$ are fed to the backbone $\mathcal{E}$, followed by emotion encoder $\mathcal{E}_\text{e}$ and corpus encoder $\mathcal{E}_\text{c}$. These two individual encoders for encoding emotion-relevant and corpus-specific features are introduced to avoid the false alignment problem caused by highly coupled features. Specifically, we first introduce a convolution neural network (CNN) based model~\cite{CNN_Pooling,DANN_ICASSP} as the backbone to extract high-level features from low-level MFCCs. 
For emotion encoder $\mathcal{E}_\text{e}$, since the emotion typically spans the short utterance and remains constant, we introduce temporal convolution network (TCN)~\cite{TCN} to effectively learn affective representations through local extraction and global integration with multi-scale receptive fields. For corpus encoder $\mathcal{E}_\text{c}$, considering that corpus-specific property of corpus is context-dependent~\cite{domain_recognition}, we use bi-directional long short term memory (Bi-LSTM)~\cite{bilstm_ser1} as the corpus encoder $\mathcal{E}_\text{c}$ for learning contextual information. 

Formally, we feed MFCCs features $\mat{X}^\text{S}$ to the encoders to encode emotion features $\mat{E}^\text{S}$ and corpus-specific features $\mat{C}^\text{S}$ from the source corpus as follows:
    \begin{align}
    \mat{E}^\text{S}=\mathcal{E}_\text{e}(\mathcal{E}(\mat{X}^\text{S})),
    \quad \mat{C}^\text{S}=\mathcal{E}_\text{c}(\mathcal{E}(\mat{X}^\text{S})).\label{eq:encoder}
    \end{align}
Similarly, $\mat{X}^\text{T}$ from the target corpus is passed through the same encoders to obtain the $\mat{E}^\text{T}$ and $\mat{C}^\text{T}$. After obtaining these features, we apply the widely-used cross-entropy loss on the $\mat{E}^\text{S}$ for discrimination of emotion-relevant features and on the  $\mat{C}=\mat{C}^\text{S} \cup \mat{C}^\text{T}$ for the corpus-specific features, which are detailed as follows:
\begin{align}
\mathcal{L}_{\mathrm {emotion}}&=-\sum_{i=1}^{n^\text{S}} {y}_i^{\text{S}} \log\varphi_\text{e}(\mat{E}_i^\text{S}),\label{eq:emotion_CE}\\
\mathcal{L}_{\mathrm {corpus}}&=-\sum_{i=1}^{n^\text{S}+n^\text{T}} d_i\log \varphi_\text{c}(\mat{C}_i)+(1-d_i)\log \left(1-\varphi_\text{c}(\mat{C}_i)\right),\label{eq:domain_CE}
\end{align}
where $\varphi_\text{e}$ and $\varphi_\text{c}$ are emotion and corpus classifiers with the softmax outputs, respectively, and $d_i$ denotes the binary corpus label for $i$-th sample, which indicates whether $C_i$ come from the source corpus or from the target corpus.

\noindent{\textbf{Contrastive decoupling loss.}}\quad
Since a prototype is very close to an representative observation in a feature space~\cite{prototype_1}, to decouple emotion and corpus features, we devise two kinds of prototypes for each corpus: emotion-relevant prototype $\mat{p}^\text{S}_{\text{e}}$ representing the center of $\mat{E}^\text{S}$ from the source corpus, and corpus-specific prototype $\mat{p}^\text{S}_{\text{c}}$ representing the center of $\mat{C}^\text{S}$ from the source corpus, which are defined as the mean representation from a projection head $g$ in a mini-batch:
\begin{align}
\mat{p}^\text{S}_{\text{e}} = \frac{1}{n^\text{S}}\sum_{i=1}^{n^\text{S}} g(\mat{E}^\text{S}_{i}),\quad
\mat{p}^\text{S}_{\text{c}} = \frac{1}{n^\text{S}}\sum_{i=1}^{n^\text{S}} g(\mat{C}^\text{S}_{i}),\label{eq:prototype}
\end{align}    
where $\mat{E}^\text{S}_{i}$ and $\mat{C}^\text{S}_{i}$ are the emotion and corpus features of $i$-th samples, and $\mat{p}^\text{T}_{\text{e}}$ and $\mat{p}^\text{T}_{\text{c}}$ can be obtained similarly. Note that prototypes are $\ell_2$-normalized.
We then can leverage these prototypes to achieve decoupling by pushing emotion-relevant prototypes away from the corpus-specific prototypes and alignment by aggregating the emotion-relevant prototypes of different corpora,
formally defined as follows:
\begin{align}
    \mathcal{L}_{\mathrm{decouple}}
    =-\frac{1}{2} \sum_{(m, n)} \log 
    \frac
    {\exp \left(\frac{\mat{p}^{m}_{\text{e}} \cdot \mat{p}^{n}_{\text{e}}}{ \tau_\text{p}}\right)}
    {\exp \left(\frac{\mat{p}^{m}_{\text{e}} \cdot \mat{p}^{m}_{\text{c}}}{ \tau_\text{p}}\right) + 
    \exp \left(\frac{\mat{p}^{m}_{\text{e}} \cdot \mat{p}^{n}_{\text{c}}}{ \tau_\text{p}}\right)},\label{eq:prototypeloss}
\end{align}
where $(m, n) \in \{(\text{S}, \text{T}), (\text{T}, \text{S})\}$, $\text{S}$ and $\text{T}$ denote the source and target corpus, respectively. $\tau_\text{p}$ is the temperature hyper-parameter of prototype-level loss, with `$\cdot$' denoting the inner product.

\subsection{Dual-Level Emotion Alignment}
\noindent{\textbf{Adaptive threshold pseudo-labeling.}}\quad 
Since the label information of the target corpus is unknown in UDA, it is difficult to maintain class consistency during corpus alignment. Inspired by~\cite{freematch}, we introduce an adaptive threshold pseudo-labeling module based on the confidence of model prediction on the target corpus to avoid class confusion while aligning corpus. Concretely, we estimate the threshold $\tau_t$ as the exponential moving average (EMA) of the confidence at each training time step $t$ (iteration) as follows:
\begin{equation}
\tau_t= \begin{cases}\frac{1}{n^C}, & \text { if } t=0 \\ \lambda \tau_{t-1}+(1-\lambda) \frac{1}{n^ \text{T}} \sum_{j=1}^{n^\text{T}} \max (\vct{q}_j), & \text { if } t>0 \end{cases}\label{eq:labeling}
\end{equation}
where $\lambda \in (0,1)$ is the momentum decay of EMA, $n^C$ is the number of classes, and $\vct{q}_j$ indicates the confidence (\ie probability) of the $j$-th sample in a mini-batch of the target corpus. The low threshold during initial training stages can involve more potentially correct samples. As the model attains greater confidence through ongoing learning, the threshold is adaptively increased. Finally, we obtain the most probable class of the target corpus predicted from the source network and then build a pseudo-labeled target corpus $\hat{\mathcal{D}}^\text{T}$ according to the threshold $\tau_t$:
\begin{align}
\hat{\mathcal{D}}^\text{T}=\left\{(\vct{X}_j^{\text{T}}, \hat{{y}}_j^{\text{T}})\mid \max (\vct{q}_j) \geq \tau_t \right\}_{j=1}^{n^\text{T}},\label{eq:pseudo-corpus}
\end{align}
where $\hat{{y}}_j^{\text{T}}$ is the pseudo label of the $j$-th sample $\vct{X}_j^{\text{T}}$. 

\noindent{\textbf{Contrastive class alignment loss.}}\quad 
The contrastive class alignment loss aims to align emotion features at the class level, which is based on the pseudo labels to maximize the mutual information of emotion features sharing the same label. Specifically, after getting the pseudo-labeled target corpus $\hat{\mathcal{D}}^\text{T}$ with $\hat{n}^\text{T}$ samples,  we treat the features $\mat{E}^\text{S}$ and $\mat{E}^\text{T}$ in $\hat{\mathcal{D}}^\text{T}$ as positive pairs if they have same emotion (pseudo) labels, and negative pairs if they have different labels.
 We then construct a new labeled data set $\mathcal{H}=\mathcal{D}^\text{S}\cup \hat{\mathcal{D}}^\text{T}$ with $n^\text{H}=n^\text{S}+\hat{n}^\text{T}$ samples. We can get the projected representation $\mat{z}_i$ of $i$-th feature $\mat{h}_i$ in $\mathcal{H}$ as follows:
\begin{align}
    \mat{z}_i = g(\mat{h}_i),
\end{align}
where $\mat{z}_i$ is $\ell_2$-normalized. Then we maximize the mutual information of representations from the positive pairs to enrich the class-discriminative representation. However, since calculating mutual information is notoriously difficult, we employs the following supervised contrastive loss (\contrastiveloss)~\cite{SupCL} as a proxy objective to maximize this mutual information:
\begin{align}
    \mathcal{L}_{\text{\contrastiveloss}}(\mathcal{D}^\text{S}, \hat{\mathcal{D}}^\text{T})
    =
    - \sum_{i=1}^{n^H} \frac{1}{\left|\mathcal{I}_i\right|} \sum_{j \in \mathcal{I}_i} \log \frac{\exp \left(\frac{\mat{z}_i \cdot \mat{z}_j}{ \tau_\text{s}} \right)}{\sum\limits_{k \neq i} \exp \left(\frac{\mat{z}_i \cdot \mat{z}_k}{ \tau_\text{s}}\right)},
\end{align}
where $\mathcal{I}_i=\{j \in\{1, \ldots, n^H\} \mid j \neq i, {y}_j={y}_i\}$ is the set of positive samples indexes to $i$-th representation $\mat{z}_i$ from $\mathcal{H}$, ${y}_i$ is the class label of $i$-th sample, and $\tau_\text{s}$ is the temperature hyper-parameter for \contrastiveloss.

\noindent{\textbf{Explicit corpus alignment loss.}}\quad 
Due to the significant domain shifts between two corpora, it is challenging to achieve class alignment with a few confident samples. To enable more efficient class alignment and learn the corpus-invariant features, we conduct the explicit corpus alignment loss at the corpus level through multiple kernels of maximum mean discrepancies (MK-MMD)~\cite{MKMMD}. Concretely, the MK-MMD is defined as the reproducing kernel Hillbert space (RKHS) $\mathcal{H}_k$ distance between the mean emotion representation of $\mat{E}^\text{S}$ and $\mat{E}^\text{T}$ from the $\mathcal{D}^\text{S}$ and $\mathcal{D}^\text{T}$ as follows:
\begin{align}
d_k^2(\mathcal{D}^\text{S}, \mathcal{D}^\text{T}) &\triangleq\left\|\mathbb{E}_P\left[\phi\left(\mat{E}^\text{S}\right)\right]-\mathbb{E}_Q\left[\phi\left(\mat{E}^\text{T}\right)\right]\right\|_{\mathcal{H}_k}^2, 
\end{align}
where $\phi(\cdot)$ denotes the kernel feature map of RKHS, and the characteristic kernel $k(\mat{E}^\text{S}, \mat{E}^\text{T})=\langle\phi(\mat{E}^\text{S}), \phi(\mat{E}^\text{T})\rangle$ associated with $\phi(\cdot)$ is defined as the linear combination of $m$ positive-definite kernels (\ie Gaussian kernels) $\left\{k=\sum_{u=1}^m \beta_u k_u\mid \sum_{u=1}^m \beta_u=1\right\}$. Here, $\beta_u$ is the weight. In this paper, we optimize $d_k^2$ according to~\cite{DAN} in practice.

Then, the dual-level emotion alignment loss $\mathcal{L}_\text{\alignmentloss}$ can guide model to learn the corpus-invariant with discrimination by combining explicit corpus alignment loss $d_k^2$ and contrastive class alignment loss $\mathcal{L}_\text{\contrastiveloss}$ as follows:
\begin{align}
\mathcal{L}_\text{\alignmentloss}(\mathcal{D}^\text{S}, \mathcal{D}^\text{T})
= d_k^2(\mathcal{D}^\text{S}, \mathcal{D}^\text{T})+\alpha_1 \vphantom{d_k^2}\mathcal{L}_{\text{\contrastiveloss}}(\mathcal{D}^\text{S}, \hat{\mathcal{D}}^\text{T}),   \label{eq:alignmentloss}
\end{align}
where $\alpha_1$ is the trade-off hyper-parameter balancing corpus alignment and class alignment. Note that $\mathcal{L}_{\text{\contrastiveloss}}$ is computed from the source corpus and pseudo-labeled target corpus.

\subsection{Objective Function for \methodname}

Overall, the proposed domain adaptation framework \methodname is illustrated in Algorithm~\ref{alg}, and the objective is defined as follows: 
\begin{align}
    \mathcal{L}_\mathrm{\methodname}= \mathcal{L}_{\mathrm {emotion}}+ \mathcal{L}_{\mathrm {corpus}}+ \mathcal{L}_\text{\alignmentloss}+ \mathcal{L}_\text{\prototypeloss}.\label{eq:allloss}
\end{align}
Note that the $\mathcal{L}_{\mathrm {emotion}}$ is computed only on the source corpus and $\mathcal{L}_{\mathrm {corpus}}$ is computed on both corpora.
\SetKwInOut{Notation}{Notation}
\begin{algorithm}[t]
% \small
%\SetAlgoLined
 \KwIn{Source $\mathcal{D}^\text{S}$; Target $\mathcal{D}^\text{T}$; Backbone $\mathcal{E}$; Emotion encoder $\mathcal{E}_\text{e}$; Corpus encoder $\mathcal{E}_\text{c}$; Projection head $g$; Emotion classifier $\varphi_\text{e}$; Corpus classifier $\varphi_\text{c}$; Mini-batch number $M$.}
 \KwOut{SER model $\varphi_\text{e} (\mathcal{E}_\text{e}(\mathcal{E}(\cdot)))$.}
 \While{training not converged}{
 \For{$t=1:M$}    
        { 
        Forward a mini-batch through the $\mathcal{E}$, $\mathcal{E}_\text{e}$, and $\mathcal{E}_\text{c}$ to get $\mat{E}^\text{S}, \mat{E}^\text{T}, \mat{C}^\text{S}$, and $\mat{C}^\text{T}$\;
        // \textbf{\emph{Classification}}: \\
        Calculate $\mathcal{L}_{\mathrm {emotion}}$ by $\varphi_\text{e} ({\mat{E}}^\text{S})$ with Eq.~\eqref{eq:emotion_CE}\;
        Calculate $\mathcal{L}_{\mathrm {corpus}}$ by $\varphi_\text{c} (\mat{{C}}^\text{S} \cup \mat{{C}}^\text{T})$ with Eq.~\eqref{eq:domain_CE}\;
        // \textbf{\emph{Contrastive emotion decoupling}}: \\
        $\mat{p}^\text{S}_{\text{e}} = \frac{1}{n^\text{S}}\sum_{i=1}^{n^\text{S}} g(\mat{E}^\text{S}_{i}),$\space $
\mat{p}^\text{S}_{\text{c}} = \frac{1}{n^\text{S}}\sum_{i=1}^{n^\text{S}} g(\mat{C}^\text{S}_{i})$\;
        $\mat{p}^\text{T}_{\text{e}} = \frac{1}{n^\text{T}}\sum_{i=1}^{n^\text{T}} g(\mat{E}^\text{T}_{i}),$\space$
\mat{p}^\text{T}_{\text{c}} = \frac{1}{n^\text{T}}\sum_{i=1}^{n^\text{T}} g(\mat{C}^\text{T}_{i})$\;
        Calculate $\mathcal{L}_\text{\prototypeloss}$ by all prototypes with Eq.~\eqref{eq:prototypeloss}\;
        % \textbf{\emph{Adaptive threshold pseudo-labeling}}: \\
        // \textbf{\emph{Dual-level emotion alignment}}: \\
        Update threshold $\tau_t$ with Eq.~\eqref{eq:labeling}\;
        $\hat{\mathcal{D}}^\text{T}=\{(\mat{X}_j^{\text{T}}, \hat{y}_j^{\text{T}})\mid \max (\vct{q}_j) \geq \tau_t \}_{j=1}^{n^\text{T}}$\;
        % $\hat{\mat{X}}_{}^\text{T} = \{\mat{X}_j^\text{T} \mid  \mat{X}_j^\text{T} \in  \hat{\mathcal{D}}^\text{T}\}_{j=1}^{n^\text{T}}$, \space$\hat{\mat{\mathcal{E}}}^\text{T}=E_\text{e}\left(E(\hat{\mat{X}}_{}^\text{T})\right)$\;
        Calculate $\mathcal{L}_\text{\alignmentloss}$ by $\mathcal{D}^\text{S}, \mathcal{D}^\text{T}$ and $\hat{\mathcal{D}}^\text{T}$ with Eq.~\eqref{eq:alignmentloss}\;
        // \textbf{\emph{Backpropagation}}: \\
        Calculate $\mathcal{L}_\mathrm{\methodname}$ with Eq.~\eqref{eq:allloss}\;
        Backpropagate $\mathcal{L}_\mathrm{\methodname}$ and update parameters.
        }
 }
 \caption{Training Procedure of \methodname}
 \label{alg}
\end{algorithm}

\section{Experiments}

\label{Experiments}
\subsection{Experimental Setup}
\noindent{\textbf{Datasets.}}\quad
To sufficiently compare the performance with state-of-the-art (SOTA) methods on the cross-corpus SER task, we conduct experiments on 6 benchmark SER corpora in different languages, including 
Chinese corpus CASIA~\cite{CASIA}, 
German corpus EMODB~\cite{EMODB}, 
Italian corpus EMOVO~\cite{EMOVO}, 
English corpora IEMOCAP~\cite{IEMOCAP}, RAVDESS~\cite{RAVDE}, and SAVEE~\cite{SAVEE}.  

For these SER datasets available, using shared fine-grained categorical emotion labels (\eg angry, happy, neutral, etc.) could harm the learning of generalizable representations due to insufficient data. Therefore, we follow the previous studies~\cite{DANN_ICASSP, ADDI_SOTA,VA_SER_1} to exploit the available data by mapping categorial emotion to dimensional emotion to aggregate the emotion information and ensure better learning of generalizable representations. Specifically, we introduce common theoretical emotional mapping Geneva Emotion Wheel~\cite{scherer2005emotions}, which is used in most works regarding emotion measurement. In this way, we define the cross-corpus SER task as two binary classification tasks by mapping the categorical emotion into to dimensional emotion, as summarized in Table~\ref{tab:rule}. 
\begin{table}[!ht]
\small
    \centering
    \caption{Mapping of emotions to two binary classification tasks, arousal and valence.}
    % \resizebox{\linewidth}{!}{
    \begin{tabular}{ccc}
    \shline
        \multirow{3}{*}{\textbf{Arousal}} & \textbf{Low} & \textbf{High} \\ \cline{2-3}
         & \makecell[c]{boredom, calm,\\neutrality, sadness}
          & \makecell[c]{anger, disgust, fear,\\happiness, surprise}  \\ \hline\hline
        \multirow{3}{*}{\textbf{Valence}} & \textbf{Negative} & \textbf{Positive} \\ \cline{2-3}
         & \makecell[c]{anger, boredom, disgust,\\fear, sadness}  & \makecell[c]{calm, happiness, \\neutrality, surprise}  \\ \shline
    \end{tabular}
    \label{tab:rule}
\end{table}
\begin{table*}[ht]
\small
    \centering
    \caption{Performance comparison (mean $\pm$ std \%) of different methods for arousal recognition. 
    For each source corpus, the results report the average WAR and F1 of its five cross-corpus  scenarios. ``Source Only'' denotes that the model is trained on the source corpus, and then is directly applied to other target corpora without any adaptation techniques. (Bold is the best)
}

    \resizebox{\linewidth}{!}{
    \begin{tabular}{r|cccccccccccc|cc}
    \shline
        Source & \multicolumn{2}{c}{CASIA} &  \multicolumn{2}{c}{EMODB} &  \multicolumn{2}{c}{EMOVO} &  \multicolumn{2}{c}{IEMOCAP} &  \multicolumn{2}{c}{RAVDESS} &  \multicolumn{2}{c}{SAVEE} &  \multicolumn{2}{c}{Average} \\ 
        \cline{1-15}
        Method & WAR & F1 & WAR & F1 & WAR & F1 & WAR & F1 & WAR & F1 & WAR & F1 & WAR\textsubscript{avg} & F1\textsubscript{avg} \\ \hline
        Source Only & 50.4\std{$\pm$0.7}& 32.4\std{$\pm$1.3}& 52.3\std{$\pm$1.6}& 49.0\std{$\pm$5.8}& 64.8\std{$\pm$0.3}& 66.1\std{$\pm$1.3}& 67.5\std{$\pm$0.3}& 79.5\std{$\pm$0.2}& 62.3\std{$\pm$1.2}& 72.2\std{$\pm$0.7}& 62.8\std{$\pm$0.5}& 75.9\std{$\pm$0.2}& 60.0\std{$\pm$0.5}& 62.5\std{$\pm$1.4} \\ \hline
        DANN~\cite{DANN_SOTA} & 60.7\std{$\pm$5.7}&	68.5\std{$\pm$9.7} & 48.9\std{$\pm$6.9}&	57.3\std{$\pm$8.5} & 56.9\std{$\pm$1.4}&	67.1\std{$\pm$3.1} & 68.9\std{$\pm$1.2}&	77.3\std{$\pm$0.7} & 64.1\std{$\pm$0.8}&	72.5\std{$\pm$1.9} & 66.2\std{$\pm$2.5}&	72.8\std{$\pm$4.2} & 61.0\std{$\pm$1.1}&	69.2\std{$\pm$1.5} \\ 
        DAN~\cite{DAN} & 67.2\std{$\pm$0.8}&	74.6\std{$\pm$1.2}&	58.6\std{$\pm$1.7}&	64.4\std{$\pm$2.1}&	61.4\std{$\pm$5.2}&	70.2\std{$\pm$5.5}&	66.7\std{$\pm$0.8}&	72.8\std{$\pm$0.8}&	65.4\std{$\pm$1.6}&	73.4\std{$\pm$1.0}&	64.1\std{$\pm$3.4}&	71.9\std{$\pm$1.8}&	63.9\std{$\pm$0.5}&	71.2\std{$\pm$0.7} \\ 
        JAN~\cite{JAN_SOTA} & 70.0\std{$\pm$2.7}& 76.2\std{$\pm$1.3}& 53.8\std{$\pm$4.4}& 59.5\std{$\pm$4.1}& 63.3\std{$\pm$3.0}& 70.7\std{$\pm$1.0}& 68.2\std{$\pm$3.2}& 74.6\std{$\pm$2.6}& 68.0\std{$\pm$3.9}& 74.8\std{$\pm$3.6}& 62.2\std{$\pm$4.5}& 69.1\std{$\pm$4.2}& 64.2\std{$\pm$1.0}& 70.8\std{$\pm$1.0} \\ 
        CDAN~\cite{CDAN_SOTA} & 63.9\std{$\pm$3.0}& 75.5\std{$\pm$2.3}& 63.4\std{$\pm$3.5}& 70.8\std{$\pm$7.5}& 67.5\std{$\pm$6.7}& 73.7\std{$\pm$7.2}& 69.2\std{$\pm$0.9}& \best{80.0}\std{$\pm$0.9}& 67.5\std{$\pm$4.3}& 77.3\std{$\pm$1.6}& 69.4\std{$\pm$4.1}& 78.0\std{$\pm$2.1}& 66.8\std{$\pm$3.4}& 75.9\std{$\pm$3.3} \\ 
        BSP~\cite{BSP_SOTA} & 67.2\std{$\pm$1.5}& 76.2\std{$\pm$1.9}& 58.5\std{$\pm$7.7}& 69.2\std{$\pm$10.3}& 59.1\std{$\pm$2.8}& 69.5\std{$\pm$5.7}& 69.8\std{$\pm$1.3}& 76.7\std{$\pm$1.5}& 72.6\std{$\pm$1.6}& 79.1\std{$\pm$1.3}& 69.0\std{$\pm$1.9}& 77.4\std{$\pm$2.2}& 66.0\std{$\pm$2.8}& 74.7\std{$\pm$3.8} \\  \hline
        CLSTM~\cite{CLSTM_SOTA} & 69.0\std{$\pm$1.8} &	73.1\std{$\pm$2.8} & 56.9\std{$\pm$0.6} &	50.0\std{$\pm$1.1} & 56.7\std{$\pm$1.1} &	60.0\std{$\pm$2.1} & 70.8\std{$\pm$0.1} &	79.1\std{$\pm$0.3} & 68.3\std{$\pm$0.3}	& 78.1\std{$\pm$0.2} & 64.8\std{$\pm$1.6} &	75.3\std{$\pm$0.8} & 64.4\std{$\pm$0.2} &	69.3\std{$\pm$0.3} \\ 
        ADDI~\cite{ADDI_SOTA} & 65.3\std{$\pm$1.7}&	66.4\std{$\pm$2.3} & 60.0\std{$\pm$3.7} &	56.0\std{$\pm$6.0} & 65.8\std{$\pm$0.4} &	71.2\std{$\pm$1.0} & 71.2\std{$\pm$1.9} &	77.2\std{$\pm$3.0} & 69.3\std{$\pm$0.8} &	75.7\std{$\pm$0.4} & 70.8\std{$\pm$0.7}	& 77.7\std{$\pm$0.3} & 67.1\std{$\pm$0.5} &	70.7\std{$\pm$1.0} \\ 
        DIFL~\cite{DANN_ICASSP} & 65.4\std{$\pm$0.5} &	74.5\std{$\pm$1.2} & 63.8\std{$\pm$3.9}	& 66.3\std{$\pm$8.1} & 61.6\std{$\pm$2.6} &	72.8\std{$\pm$4.9} & 71.7\std{$\pm$1.5} &	79.4\std{$\pm$0.6} & 68.4\std{$\pm$0.2} &	77.5\std{$\pm$0.6} & 68.5\std{$\pm$3.2}	& 76.3\std{$\pm$4.2} & 66.6\std{$\pm$1.4} &	74.5\std{$\pm$2.3} \\ 
        CAAM~\cite{CPAC_IJCAI} & 66.5\std{$\pm$4.5} &	75.3\std{$\pm$2.9}& 68.1\std{$\pm$0.7}&	73.6\std{$\pm$0.3} & 65.8\std{$\pm$1.0} &	75.3\std{$\pm$0.7} & 68.9\std{$\pm$1.7} &	79.1\std{$\pm$0.5} & 69.7\std{$\pm$2.5} &	75.2\std{$\pm$2.9} & 70.5\std{$\pm$0.5}	& 79.0\std{$\pm$0.3} & 68.3\std{$\pm$0.3} &	76.2\std{$\pm$0.3} \\ \hline
        \textbf{\methodname} & \best{73.8}\std{$\pm$3.2}	& \best{79.9}\std{$\pm$2.1} & \best{72.7}\std{$\pm$0.5}	& \best{78.2}\std{$\pm$0.9} & \best{77.0}\std{$\pm$1.3} &	\best{82.3}\std{$\pm$0.9} & \best{75.1}\std{$\pm$1.1}	 & 79.8\std{$\pm$1.3} & \best{73.5}\std{$\pm$0.8}	& \best{80.0}\std{$\pm$0.4} & \best{73.9}\std{$\pm$1.2}	& \best{79.4}\std{$\pm$0.9} & \best{74.3}\std{$\pm$0.3}	& \best{79.9}\std{$\pm$0.4} \\ 
        \shline
    \end{tabular}
    }
    \label{tab:sota_arousal}
\end{table*}

\noindent{\textbf{Preprocessing.}}\quad
We utilize Librosa toolbox~\cite{librosa} to extract the most commonly-used Mel-Frequency Cepstral Coefficients (MFCCs) features~\cite{GM_TCNet,LIGHTSER,TIMNET_SER} as the inputs to \methodname. We extract the first 40 coefficients to obtain the low-frequency envelope and high-frequency details in this paper.

\noindent{\textbf{Task setup.}}\quad For the cross-corpus SER tasks, 6 corpora form 30 cross-corpus SER scenarios; \eg a scenario like CASIA$\rightarrow$EMODB indicates the model is trained with labeled corpus CASIA and unlabeled corpus EMODB, and then be tested on EMODB.

\noindent{\textbf{Evaluation metrics.}}\quad 
We use two widely-used metrics for the cross-corpus SER task: weighted accuracy (WAR) and F1-score (F1). WAR uses the class probabilities to balance the recall metric of different classes, and F1-score is a metric that combines precision and recall by taking their harmonic mean.

\noindent{\textbf{Baselines.}}\quad
To prove the effectiveness of our method, we compare our method with the following methods: the baselines for unsupervised domain adaptation methods are DANN~\cite{DANN_SOTA}, DAN~\cite{DAN}, JAN~\cite{JAN_SOTA}, CDAN~\cite{CDAN_SOTA}, and BSP~\cite{BSP_SOTA}; the cross-corpus SER baseline methods are CLSTM~\cite{CLSTM_SOTA}, ADDI~\cite{ADDI_SOTA}, DIFL~\cite{DANN_ICASSP}, and CAAM~\cite{CPAC_IJCAI}.

\noindent{\textbf{Implementation details.}}\quad
Our experiments are performed on the PyTorch platform and each experiment is repeated 3 times to show the robustness of the methods. For the proposed method, the backbone $\mathcal{E}$ has 3 local feature learning blocks, each of which consists of a 1D convolution operation (with kernel size 3 and output channel size 64) followed by batch normalization, \textit{ReLU} activation function, dropout (dropout rate is 0.1) and a max pooling layer. Then the emotion encoder $\mathcal{E}_\text{e}$ is composed of 5 residual blocks of temporal convolution network~\cite{TCN}. We set the kernel size 2 and replace the weight normalization with the batch normalization. We use Bi-LSTM~\cite{bilstm_ser1} as the $\mathcal{E}_\text{c}$ with 64 hidden units. The temperature hyper-parameters $\tau_\text{p}$ and $\tau_\text{s}$ are both set to 1e-2. The trade-off hyper-parameter $\alpha_1$ is 0.1. In addition, we adopt RAdam~\cite{RAdam} to optimize the model with a learning rate 1e-3, and a mini-batch size $M$ of 64. 
For the UDA baselines, we use the same inputs and encoder (\ie $\mathcal{E}+\mathcal{E}_\text{e}$) to extract emotion features for a fair comparison. We reproduce all cross-corpus SER baselines following their original paper and default settings.

\begin{table*}[ht]
\small
    \centering
    \caption{Performance comparison (mean $\pm$ std \%) of different methods for valence recognition. For each source corpus, the results report the average WAR and F1 of its five cross-corpus  scenarios. ``Source Only'' denotes that the model is trained on the source corpus, and then is directly applied to other target corpora without any adaptation techniques. (Bold is the best)
    }
    \resizebox{\linewidth}{!}{
    \begin{tabular}{r|cccccccccccc|cc}
    \shline
        Source & \multicolumn{2}{c}{CASIA} &  \multicolumn{2}{c}{EMODB} &  \multicolumn{2}{c}{EMOVO} &  \multicolumn{2}{c}{IEMOCAP} &  \multicolumn{2}{c}{RAVDESS} &  \multicolumn{2}{c}{SAVEE} &  \multicolumn{2}{c}{Average} \\ 
        \cline{1-15}
        Method & WAR & F1 & WAR & F1 & WAR & F1 & WAR & F1 & WAR & F1 & WAR & F1 & WAR\textsubscript{avg} & F1\textsubscript{avg} \\ \hline
        Source Only & 48.6\std{$\pm$1.5}& 31.9\std{$\pm$1.5}& 51.9\std{$\pm$0.4}& 13.4\std{$\pm$0.9}& 52.6\std{$\pm$1.5}& 13.0\std{$\pm$3.5}& 55.4\std{$\pm$0.3}& 26.2\std{$\pm$3.6}& 57.0\std{$\pm$0.5}& 51.2\std{$\pm$4.1}& 47.3\std{$\pm$1.1}& 40.4\std{$\pm$2.0}& 52.1\std{$\pm$0.0}& 29.4\std{$\pm$0.3} \\ \hline
        DANN~\cite{DANN_SOTA} & 53.6\std{$\pm$1.0}&	49.4\std{$\pm$6.1} & 52.5\std{$\pm$4.6}&	41.8\std{$\pm$13.0} & 53.1\std{$\pm$1.7}&	42.0\std{$\pm$3.8} & 54.9\std{$\pm$1.7}&	58.2\std{$\pm$2.5}  & 53.8\std{$\pm$2.2}&	50.4\std{$\pm$3.8} & 47.2\std{$\pm$1.6}&	44.8\std{$\pm$2.2} & 52.5\std{$\pm$1.3}&	47.8\std{$\pm$1.4} \\ 
        DAN~\cite{DAN} & 54.7\std{$\pm$1.7}& 52.3\std{$\pm$0.2}& 54.6\std{$\pm$0.4}& 39.6\std{$\pm$1.1}& 56.8\std{$\pm$1.2}& 51.6\std{$\pm$1.4}& 53.8\std{$\pm$1.8}& 57.9\std{$\pm$0.9}& 56.0\std{$\pm$0.5}& 55.1\std{$\pm$2.0}& 54.3\std{$\pm$1.7}& 54.5\std{$\pm$1.9}& 55.0\std{$\pm$0.3}& 51.8\std{$\pm$0.3} \\ 
        JAN~\cite{JAN_SOTA} & 57.0\std{$\pm$1.0}& 57.2\std{$\pm$1.7}& 54.2\std{$\pm$1.4}& 49.2\std{$\pm$4.0}& 54.1\std{$\pm$1.7}& 52.2\std{$\pm$3.7}& 54.6\std{$\pm$0.7}& 59.2\std{$\pm$1.9}& 58.4\std{$\pm$1.6}& 57.4\std{$\pm$3.2}& 53.9\std{$\pm$1.4}& 52.6\std{$\pm$3.3}& 55.4\std{$\pm$0.6}& 54.6\std{$\pm$0.3} \\ 
        CDAN~\cite{CDAN_SOTA} & 53.9\std{$\pm$1.1}& \best{63.1}\std{$\pm$0.8}& 55.4\std{$\pm$2.1}& 62.0\std{$\pm$6.6}& 55.4\std{$\pm$2.6}& 53.1\std{$\pm$9.7}& \best{56.0}\std{$\pm$1.1}& 55.6\std{$\pm$2.6}& 57.5\std{$\pm$2.9}& 52.6\std{$\pm$9.5}& 53.9\std{$\pm$0.3}& 49.4\std{$\pm$6.3}& 55.3\std{$\pm$0.4}& 56.0\std{$\pm$5.0} \\ 
        BSP~\cite{BSP_SOTA} & 52.2\std{$\pm$1.5}& 55.8\std{$\pm$5.0}& 56.2\std{$\pm$1.7}& 56.0\std{$\pm$4.6}& 52.8\std{$\pm$0.5}& 50.2\std{$\pm$6.9}& 55.3\std{$\pm$2.2}& 56.2\std{$\pm$1.8}& 56.4\std{$\pm$0.9}& 54.5\std{$\pm$3.9}& 53.9\std{$\pm$1.1}& 51.4\std{$\pm$7.8}& 54.5\std{$\pm$0.6}& 54.0\std{$\pm$4.4} \\ \hline
        CLSTM~\cite{CLSTM_SOTA} & 57.7\std{$\pm$1.6}&	56.8\std{$\pm$0.4} & 54.2\std{$\pm$0.9}&	60.0\std{$\pm$4.0} & 51.4\std{$\pm$0.4}&	41.8\std{$\pm$4.2} & 53.8\std{$\pm$0.1}&	\best{61.4}\std{$\pm$0.4} & 56.6\std{$\pm$0.3}&	57.1\std{$\pm$0.7} & 56.9\std{$\pm$0.5}&	51.3\std{$\pm$5.2} & 55.1\std{$\pm$0.3}&	54.7\std{$\pm$0.7} \\ 
        ADDI~\cite{ADDI_SOTA} & 57.7\std{$\pm$1.3}	&62.2\std{$\pm$1.3} & 55.3\std{$\pm$0.6}	&\best{65.0}\std{$\pm$1.3} & 52.4\std{$\pm$2.3}	&51.3\std{$\pm$0.5} & 55.9\std{$\pm$0.3}	&58.9\std{$\pm$0.5} & 58.1\std{$\pm$1.2}	&56.8\std{$\pm$1.8} & 55.3\std{$\pm$2.2}	&50.0\std{$\pm$5.2} & 55.8\std{$\pm$0.6}	&57.4\std{$\pm$0.9} \\ 
        DIFL~\cite{DANN_ICASSP} & 52.6\std{$\pm$0.4}	& 60.1\std{$\pm$1.5} & 54.4\std{$\pm$3.3}	& 52.3\std{$\pm$8.2} & 56.1\std{$\pm$0.3}	& 50.7\std{$\pm$3.5} & 44.4\std{$\pm$1.3}	& 60.8\std{$\pm$1.1} & 53.8\std{$\pm$2.0}	& 56.2\std{$\pm$7.3} & 52.5\std{$\pm$4.6}	& 54.9\std{$\pm$2.6} & 52.3\std{$\pm$0.5}	& 55.8\std{$\pm$1.0} \\ 
        CAAM~\cite{CPAC_IJCAI} & 55.0\std{$\pm$1.5}	& 57.3\std{$\pm$2.1} & 56.6\std{$\pm$1.8}	& 45.5\std{$\pm$4.0} & 54.2\std{$\pm$2.0}	& 54.8\std{$\pm$3.9} & 54.2\std{$\pm$1.3}	& 60.5\std{$\pm$1.6} & 50.1\std{$\pm$4.4}	& 51.5\std{$\pm$7.0} & 52.3\std{$\pm$2.3}	&\best{58.7}\std{$\pm$2.4} & 53.7\std{$\pm$0.5}	& 54.7\std{$\pm$1.7} \\ \hline
        \textbf{\methodname} & \best{60.2}\std{$\pm$1.0}& 59.2\std{$\pm$1.3}& \best{56.9}\std{$\pm$0.2}& 53.9\std{$\pm$2.3}& \best{61.1}\std{$\pm$1.9}& \best{57.0}\std{$\pm$1.9}& 55.9\std{$\pm$1.3}& 58.8\std{$\pm$0.9}& \best{59.5}\std{$\pm$0.4}& \best{58.0}\std{$\pm$1.4}& \best{58.1}\std{$\pm$0.8}& 58.3\std{$\pm$2.6}& \best{58.6}\std{$\pm$0.2}& \best{57.5}\std{$\pm$0.2} \\ 
        \shline
    \end{tabular}
    }
    \label{tab:sota_valence}
\end{table*}
\begin{figure*}[t]
\setlength{\abovecaptionskip}{0.1cm}
	\centering
	\includegraphics[width=0.96\linewidth]{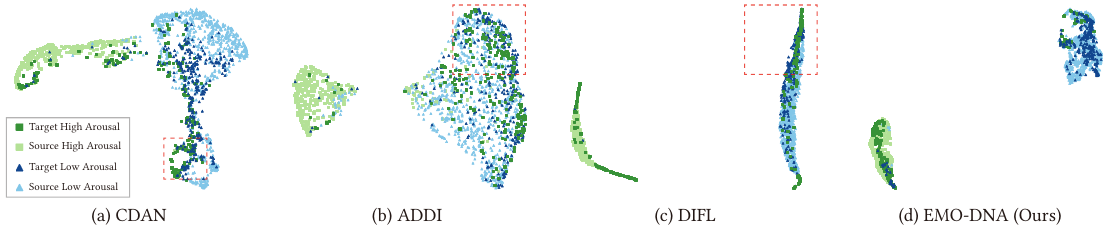}
	\caption{Feature visualizations by UMAP. The learned features by different SOTA methods under the adaptation scenario CASIA$\rightarrow$EMODB. The dashed red box is the highlighted part suffering from false alignment or class confusion problems. }
 \label{fig:sota}
\end{figure*}

\subsection{Comparison with State-of-the-Arts}
\noindent{\textbf{Quantitative comparison.}}\quad
We compare the proposed \methodname with current commonly used UDA methods and cross-corpus SER SOTA methods. In the experiment, we take each corpus as the source corpus and each of the other five corpora as the target corpus respectively, and we report the average performance of the five cross-corpus scenarios for each source corpus. 
Moreover, we show the mean and standard deviation of the results of three runs. The results are reported in Tables~\ref{tab:sota_arousal} and~\ref{tab:sota_valence}. Overall, \methodname achieves the best performance on almost all cross-corpus adaptation scenarios against the SOTA methods and gains consistent improvements on each scenario. 

For arousal recognition, compared with methods that learn corpus-invariant features, \methodname outperforms DANN, DAN, JAN, ADDI, DIFL, and CAAM by 13.3\%, 9.5\%, 9.3\%, 7.2\%, 7.7\%, and 6.0\% WAR on average, respectively. 
The inferior performance of these methods may be attributed to their tendency to overlook class consistency while focusing solely on aligning the distributions of the source and target corpora. 
Note that although DIFL additionally utilizes labels of corpus, gender and language to learn corpus-invariant features, but fails to align ambiguous target samples with correct classes due to the lack of class-discriminative information, leading to performance degradation. 

In addition, compared with methods that learn class-discriminative corpus-invariant features, \methodname outperforms CDAN and BSP by 7.5\% and 8.3\% WAR on average. 
We believe that they may suffer from the false alignment problem and cause a performance drop.

For valence recognition, all baselines show relatively poor performance. It indicates that corpus-invariant feature learning is more challenging to achieve for valence than for arousal, as concluded in~\cite{valence_worse, DANN_ICASSP}.
However, \methodname outperforms almost all baselines for valence recognition by alleviating the false alignment and class confusion problems. 

Overall, \methodname achieves greater performance improvements on both arousal and valence recognition scenarios by exploiting the emotion decoupling and alignment to learn emotion-relevant corpus-invariant features.

\noindent{\textbf{Feature visualization.}}\quad
To give an intuitive understanding of \methodname, we visualize the features of a cross-corpus adaptation scenario \emph{CASIA} $\rightarrow$ \emph{EMODB} through the popular uniform manifold approximation and projection (UMAP) technique~\cite{UMAP}. We visualize the features learned by CDAN, ADDI, and DIFL and the proposed \methodname in Figure~\ref{fig:sota} for comparison. Both ADDI and DIFL can learn good corpus-invariant features but fail to capture discriminative information, suffering from class confusion problem. 
For instance, their target high arousal features are aligned with the low arousal features rather than with the correct ones. 
On the contrary, the target features learned by \methodname demonstrate higher intra-class compactness and much larger inter-class margin. These suggest that our \methodname produces more emotion-relevant corpus-invariant features for the target corpus and substantiates our improvement shown in Tables~\ref{tab:sota_arousal} and~\ref{tab:sota_valence}.

\begin{table}[t]
\small
\centering
\caption{Ablation experiments on the 30 cross-corpus adaptation scenarios. `w/o' denotes removing the component.}
\begin{tabular*}{1\linewidth}{@{\extracolsep{\fill}}l|cccc}
\shline
 \multirow{2}{*}{Method}  &  \multicolumn{2}{c}{Arousal} &  \multicolumn{2}{c}{Valence} \\

~ & \text{WAR}\textsubscript{avg} & \text{F1}\textsubscript{avg}  &  \text{WAR}\textsubscript{avg} & \text{F1}\textsubscript{avg}          \\\hline
V1: Source-only    &      60.0\std{$\pm$0.5}& 62.5\std{$\pm$1.4}  &     52.1\std{$\pm$0.0}& 29.4\std{$\pm$0.3}      \\\hline
V2: w/o $\mathcal{L}_\text{\prototypeloss}+\mathcal{L}_\text{\alignmentloss}$   &   60.6\std{$\pm$0.1}&	66.0\std{$\pm$0.1} &    54.1\std{$\pm$0.2}&	49.7\std{$\pm$0.2}  \\
V3: w/o $\mathcal{L}_\text{\prototypeloss}$ &    71.5\std{$\pm$0.7}&	77.7\std{$\pm$0.5}  &      56.8\std{$\pm$0.2}&	55.8\std{$\pm$0.3}   \\
V4: w/o $\mathcal{L}_\text{\alignmentloss}$   &   64.4\std{$\pm$0.1}&	69.8\std{$\pm$0.2}  &     56.4\std{$\pm$0.2}&	56.8\std{$\pm$0.2}   \\ \hline
V5: w/o Decoupling &      65.1\std{$\pm$0.2}&	72.1\std{$\pm$0.1}  &      55.1\std{$\pm$0.3}&	50.6\std{$\pm$0.4}    \\\hline
\textbf{\methodname} & \best{74.3}\std{$\pm$0.3}	& \best{79.9}\std{$\pm$0.4}  &     \best{58.6}\std{$\pm$0.2}	& \best{57.5}\std{$\pm$0.2} \\
\shline
\end{tabular*}
\label{tab:ablation}
\end{table}

\subsection{Ablation Studies}
\noindent{\textbf{Ablation study on each component.}}\quad
To validate the effectiveness of each component in \methodname,  we conduct ablation studies on these 30 cross-corpus adaptation scenarios of arousal and valence recognition. We construct the five variations of \methodname for comparison. 
As shown in Table~\ref{tab:ablation}, each result denotes the average performance over the 30 cross-corpus cases, and we can get the following conclusions: 
\textit{First}, we only introduce the corpus encoder, whose results are reported in V2. From the comparison of V1 and V2, V2 achieves better results than V1 attributed to introduce an auxiliary task (\ie corpus recognition) for learning more robust representations; 
\textit{Second}, by removing the dual-level emotion alignment method, V3 cannot perceive the class discrimination information and the ambiguous target samples may be aligned with the wrong classes, thus leading to a degradation in the performance of the model; 
\textit{Third}, if contrastive emotion decoupling is ignored, the feature distributions learned in V4 and V5 may suffer from the false alignment problem even if they are well-aligned, which makes a sub-optimal adapted model;  
\textit{Last}, \methodname uses above two methods to optimize the model jointly and achieves optimal results.
\begin{figure*}[ht]
\setlength{\abovecaptionskip}{0.1cm}
	\centering
	\includegraphics[width=1.0\linewidth]{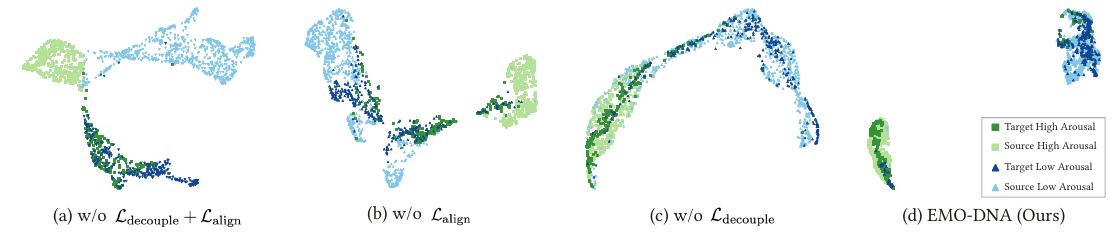}
	\caption{The learned features with different variants of \methodname under the adaptation scenario CASIA$\rightarrow$EMODB.}
 \label{fig:ablation}
\end{figure*}

\noindent{\textbf{Visualization of features with different components.}}\quad
To investigate the effects of each proposed component, we also visualize the features learned by different variants of \methodname in Figure~\ref{fig:ablation} and draw the following intuitive observations: 
(1) Figure~\ref{fig:ablation}(a) shows the features learned without the guidance of $\mathcal{L}_\text{\prototypeloss}$ and $\mathcal{L}_\text{\alignmentloss}$, where we can observe distinct class discriminability on the source corpus, severe class confusion on the target corpus, and clear domain-shifts between two corpora. This demonstrates the effectiveness of $\mathcal{L}_\text{\prototypeloss}$ and $\mathcal{L}_\text{\alignmentloss}$ in joint learning corpus-invariant and class-discriminative features.
(2) Figure~\ref{fig:ablation}(b) shows that without $\mathcal{L}_\text{\alignmentloss}$, the target and source distributions start to get closer but are not well aligned and still have the domain shifts. It indicates that feature decoupling without alignment can not produce generalized representation. 
(3) Figure~\ref{fig:ablation}(c) shows that although the $\mathcal{L}_\text{\alignmentloss}$ can align features across corpora and enhance the discriminability on the target corpus, there are still some clusters of mixed emotions on both source and target corpora. It suggests that $\mathcal{L}_\text{\prototypeloss}$ (\ie contrastive emotion decoupling) can provide decoupled emotion-relevant features for the corpus alignment to avoid the false alignment problem. 
(4) Figure~\ref{fig:ablation}(d) shows that, combined with all the proposed components, the learned distributions achieve significant inter-class margin with high intra-class compactness. Therefore, the emotion feature decoupling and alignment exhibit mutually incremental effects and jointly enable the model to achieve robust adaptability in the target corpus.

\subsection{Model Analysis}

\noindent{\textbf{Effectiveness of contrastive class alignment.}}\quad
The contrastive class alignment loss $\mathcal{L}_\text{\contrastiveloss}$ is proposed for aligning emotion features at the class level, which can be easily integrated into other UDA methods as a plug-and-play approach. To assess its efficacy, we incorporate $\mathcal{L}_\text{\contrastiveloss}$ into several typical UDA baselines. As shown in Table~\ref{tab:SCL}, DANN, DAN, and JAN trained with $\mathcal{L}_\text{\contrastiveloss}$ demonstrate substantial performance gains and improved generalizability. These methods concentrate on learning corpus-invariant features while overlooking class consistency, whereas the contrastive class alignment facilitates perceiving class discrimination information and achieving corpus-invariant class-discriminative features.
However, we notice that CDAN and BSP with $\mathcal{L}_\text{\contrastiveloss}$ show a degraded performance on arousal recognition. This could potentially be attributed to their emphasis on learning class-discriminative information that may be disrupted by $\mathcal{L}_\text{\contrastiveloss}$. Thus it may be more appropriate to combine $\mathcal{L}_\text{\contrastiveloss}$ with the corpus-invariant methods than those likewise for class discrimination. 
Overall, our proposed contrastive class alignment effectively enhances the ability of corpus-invariant UDA methods to learn class discrimination information.

\begin{table}[t]
\small
\centering
\caption{Experimental results of applying $\mathcal{L}_\text{\contrastiveloss}$ to other UDA baselines. ``$\uparrow$'' denotes a better result compared to the method without $\mathcal{L}_\text{\contrastiveloss}$ and ``$\downarrow$'' denotes a worse result.}
\begin{tabular}{r|ccccc}
\shline
 \multirow{2}{*}{\textbf{Method}}  &  \multicolumn{2}{c}{\textbf{Arousal}} &&  \multicolumn{2}{c}{\textbf{Valence}} \\
\cline{2-3}\cline{5-6}
~ & \text{WAR}\textsubscript{avg} & \text{F1}\textsubscript{avg}  &&  \text{WAR}\textsubscript{avg} & \text{F1}\textsubscript{avg}          \\\hline
DANN + $\mathcal{L}_\text{\contrastiveloss}$ &      64.1\std{$\pm$0.8}$\uparrow$&	72.4\std{$\pm$0.7}$\uparrow$  &&     55.7\std{$\pm$0.6}$\uparrow$&	57.5\std{$\pm$0.8}$\uparrow$      \\
DAN + $\mathcal{L}_\text{\contrastiveloss}$ &      65.1\std{$\pm$0.4}$\uparrow$&	72.2\std{$\pm$0.1}$\uparrow$  &&      55.8\std{$\pm$0.4}$\uparrow$&	53.7\std{$\pm$0.6}$\uparrow$    \\
JAN+ $\mathcal{L}_\text{\contrastiveloss}$  &    68.2\std{$\pm$0.6}$\uparrow$&	74.7\std{$\pm$1.8}$\uparrow$ &&    56.5\std{$\pm$0.5}$\uparrow$&	57.0\std{$\pm$0.4}$\uparrow$  \\
CDAN + $\mathcal{L}_\text{\contrastiveloss}$  &    63.5\std{$\pm$0.2}$\downarrow$&	74.5\std{$\pm$0.6}$\downarrow$ &&      56.1\std{$\pm$0.8}$\uparrow$& 58.2\std{$\pm$0.7}$\uparrow$   \\
BSP+ $\mathcal{L}_\text{\contrastiveloss}$   &   64.3\std{$\pm$0.3}$\downarrow$& 74.2\std{$\pm$1.2}$\downarrow$  &&     55.7\std{$\pm$0.3}$\uparrow$& 57.9\std{$\pm$0.5}$\uparrow$   \\
\shline
\end{tabular}
\label{tab:SCL}
\end{table}

\begin{table}[t]
\small
\centering
\caption{Ablation experiments for adaptive threshold pseudo-labeling  on the 30 cross-corpus adaptation scenarios.}
\begin{tabular}{l|ccccc}
\shline
 \multirow{2}{*}{\textbf{Method}}  &  \multicolumn{2}{c}{\textbf{Arousal}} &&  \multicolumn{2}{c}{\textbf{Valence}} \\
\cline{2-3}\cline{5-6}
~ & \text{WAR}\textsubscript{avg} & \text{F1}\textsubscript{avg}  &&  \text{WAR}\textsubscript{avg} & \text{F1}\textsubscript{avg}          \\\hline
w/o $\mathcal{L}_\text{\contrastiveloss}$ &      64.8\std{$\pm$0.4}&	70.5\std{$\pm$0.7}  && 54.1\std{$\pm$0.6}&	49.3\std{$\pm$0.8}     \\\hline
w/o threshold &      63.3\std{$\pm$0.2}&	72.0\std{$\pm$0.3}&&     53.7\std{$\pm$0.4}&	44.7\std{$\pm$0.2}     \\
Fixed threshold &      67.5\std{$\pm$3.2}&	76.4\std{$\pm$2.3}&&     55.5\std{$\pm$0.2}&	\best{58.3}\std{$\pm$0.1}      \\\hline
Adaptive threshold & \best{74.3}\std{$\pm$0.3}	& \best{79.9}\std{$\pm$0.4}  &&     \best{58.5}\std{$\pm$0.2}	& 57.4\std{$\pm$0.1} \\
\shline
\end{tabular}
\label{tab:FA}
\end{table}

\noindent{\textbf{Effectiveness of pseudo-labeling module.}}\quad
The efficacy of our contrastive class alignment loss $\mathcal{L}_\text{\contrastiveloss}$ relies on the quality of the target pseudo-labels, as using noisy pseudo-labels can exacerbate class confusion issues. Table~\ref{tab:FA} shows that when directly applying $\mathcal{L}_\text{\contrastiveloss}$ without the proposed threshold (``w/o threshold''), performance even declines compared to \methodname without the proposed loss (``w/o $\mathcal{L}_\text{\contrastiveloss}$''), which indicates that a threshold is necessary for pseudo-labeling. Additionally, setting a fixed threshold of 0.9 produces worse results, as a high threshold may cause the model to disregard potentially correct samples during early training stages, and a fixed threshold is also unsuitable for flexible cross-corpus adaptation scenarios. Hence, we conclude that the adaptive threshold pseudo-labeling module enhances the robustness of \methodname to class alignment.

\section{Conclusion}
\label{Conclusion}
In this paper, we propose \methodname, a novel emotion decoupling and alignment learning framework to learn emotion-relevant corpus-invariant features for cross-corpus SER. 
The proposed contrastive emotion decoupling module decouple emotion-relevant features from corpus-specific features in a latent distorted manifold. The proposed dual-level emotion alignment module performs corpus-level alignment and class-level alignment for effective alignment between source and target corpora. The decoupling and alignment promote progressive learning with each other and enable the model to achieve robust adaptability in the target corpus. 
Experimental results demonstrate the superiority of \methodname over state-of-the-art methods in several adaptation scenarios across six widely-use corpora. Ablation studies and visualization further validate the advantages of the synergy between decoupling and alignment. 
However, there are still some limitations in the existing experimental setup. Because most of the current SER corpora lack dimensional emotion labels, it is a rough approach to map categorical emotion to dimensional one.  
In the future, we plan to incorporate more modal information for affective analysis and make an attempt to mitigate the gap in mapping categorical emotions into dimensional one.

\clearpage
{
\bibliographystyle{ACM-Reference-Format}
\balance

}

\end{document}